# Attraction between Topological Defects in Graphene

A.I. Podlivaev, L.A. Openov

National Research Nuclear University "MEPhI", Kashirskoe sh. 31, Moscow 115409, Russian Federation

E-mail addresses: LAOpenov@mephi.ru, laopenov@gmail.com

**ABSTRACT**

The interaction of Stone–Wales topological defects in graphene has been studied through computer simulation. This simulation has revealed configurations of two defects with energies below the energy of a monolayer with two spaced defects. This indicates the attraction between defects and the possibility of the formation of their clusters. The attraction is due to the interference between defect-induced wavy distortions of the structure of the monolayer. In this case, the amplitude of transverse displacement of atoms near a pair of defects reaches 2-3 Å. Such a strong deformation of graphene by Stone–Wales defects can be one of the reasons for its experimentally observed "crumpled" texture.

Many properties of solids (strength, thermal, electric, etc.) are very sensitive to defects in their structure. In bulk crystals, elementary intrinsic defects are vacancies and interstitials. Quasi-two-dimensional materials (e.g., graphene [1]), as well as fullerenes and nanotubes, can contain so-called topological defects, which appear because of the corresponding rearrangement of interatomic bonds without removal and addition of atoms. The simplest of these defects is the Stone–Wales defect [2]. It is formed when one of the C–C bonds (core of a defect) rotates at 90°; as a result, four hexagons are transformed into two heptagons and two pentagons (see Fig. 1a). Stone-Wales defects significantly affect the structural and electrical characteristics of graphene [3]. In particular, they break the symmetry of a crystal lattice, making two sublattices nonequivalent and, thus, leading to the appearance of a band gap in the electronic spectrum [4], which is important for applications in nanoelectronics. Furthermore, Stone-Wales defects are of interest as centers of preferable adsorption of various elements [5].

A Stone-Wales defect formed in graphene does not remain planar [6] (as was erroneously thought for a long time): two atoms connected by the rotating C-C bond are displaced perpendicularly to the plane of the monolayer by ≈ 0.3 Å in opposite directions, carrying a large number of other atoms located at distances up to several nanometers from the core. The difference between transverse displacements of atoms reaches 1.7 Å in a 260-atom supercell [6]. As a result, the energy of the system decreases by ≈ 0.8 eV [6, 7] and sinelike distortion of the monolayer appears (Fig. 1b). If both atoms are displaced in the same direction by ≈ 0.5 Å, the distortion of the monolayer has a cosinelike shape (Fig. 1c), which corresponds to a saddle point on the surface of the potential energy [6]. In both cases, the profile of transverse displacements of atoms has a wavy shape (transverse displacement of atoms is sign alternating) with the length close to the dimension of the supercell in the corresponding direction, and the maximum displacements are reached at a distance of ≈ 5 Å from the core.

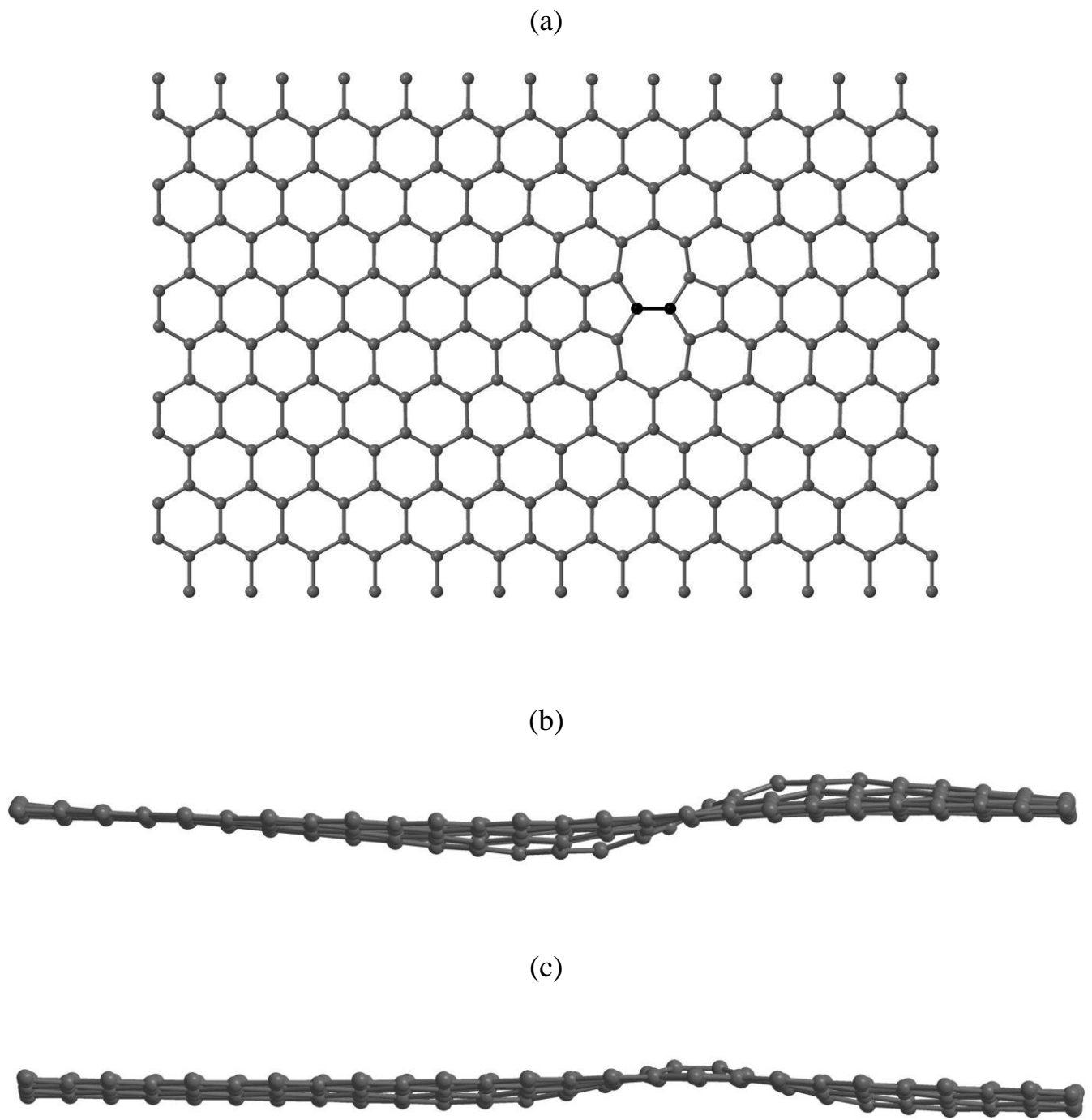

Fig. 1. Stone-Wales defect in graphene. (a) Planar Stone-Wales defect. Plan view. The C-C bond rotated by an angle of 90° (core of a defect) is indicated in black. (b) Sinelike and (c) cosinelike Stone-Wales defects in graphene (side view).

For applications, it is necessary to know, first, how the electronic characteristics and adsorptivity of graphene depends on the concentration of Stone-Wales defects [4, 5] and, second, whether Stone-Wales defects (produced, e.g., by electron irradiation) remain isolated or they can form clusters. The answer to the latter question depends on the character of the interaction between Stone–Wales defects. Ab initio calculation of the energy $E_f$ of formation of an ordered system of Stone-Wales defects for various distances $d$ between defects was performed in [8]. An increase in $E_f$ with a decrease in $d$ was interpreted as repulsion between two defects. However, the defect graphene monolayer was simulated in [8] with the use of periodic boundary conditions for a small (less than 100 atoms) supercell containing only one Stone-Wales defect. In this case, the value of $d$ is determined solely by the dimension of the supercell, and interaction occurs between the defect and its images in other supercells; i.e., the defect really interacts with itself. In addition, all effects associated with possible anisotropy of the interaction are smoothed.

In this work, the energies $E(2)$ of fairly large 252- and 260-atom supercells each containing two Stone-Wales defects are compared to the energies of defectless supercells, $E(0)$, and supercells with one defect, $E(1)$. The energy of interaction between defects is defined as $E_{int}=E(2)+E(0)-2E(1)$, i.e., by the standard formula for the binding energy of two particles [9] (positive and negative $E_{int}$ values correspond to repulsion and attraction, respectively). Periodic boundary conditions were imposed in both directions in the plane of the monolayer. The distance between defects was smaller than the distance to their images in neighboring supercells; i.e., the main contribution to $E_{int}$ came from the interaction between defects, rather than from the interaction of defects with their images. The interatomic interactions were described within the nonorthogonal tight-binding model [10], which is less accurate than ab initio approaches but gives interatomic distances and binding energies of various carbon structures in good agreement with experimental data and ab initio calculations [10, 11]. Furthermore, this model requires much less computer resources and allows the simulation of a system of several hundred atoms.

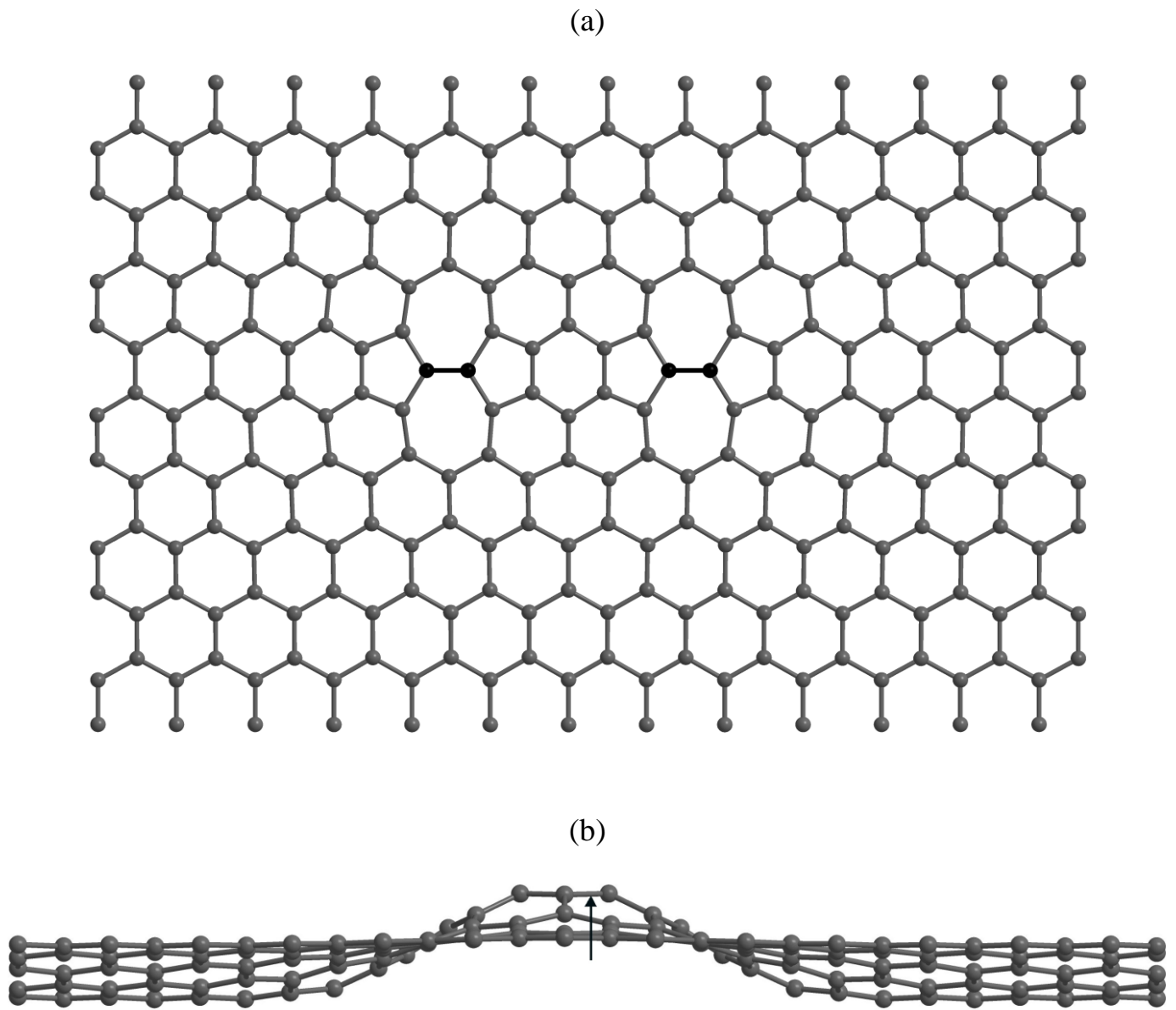

Fig. 2. Two Stone-Wales defects in configuration I. (a) Plan view. Cores of defects are shown in black. (b) Side view in the direction perpendicular to the cores of defects. The arrow indicates the maximum transverse displacements of atoms over the plane of a monolayer.

(a)

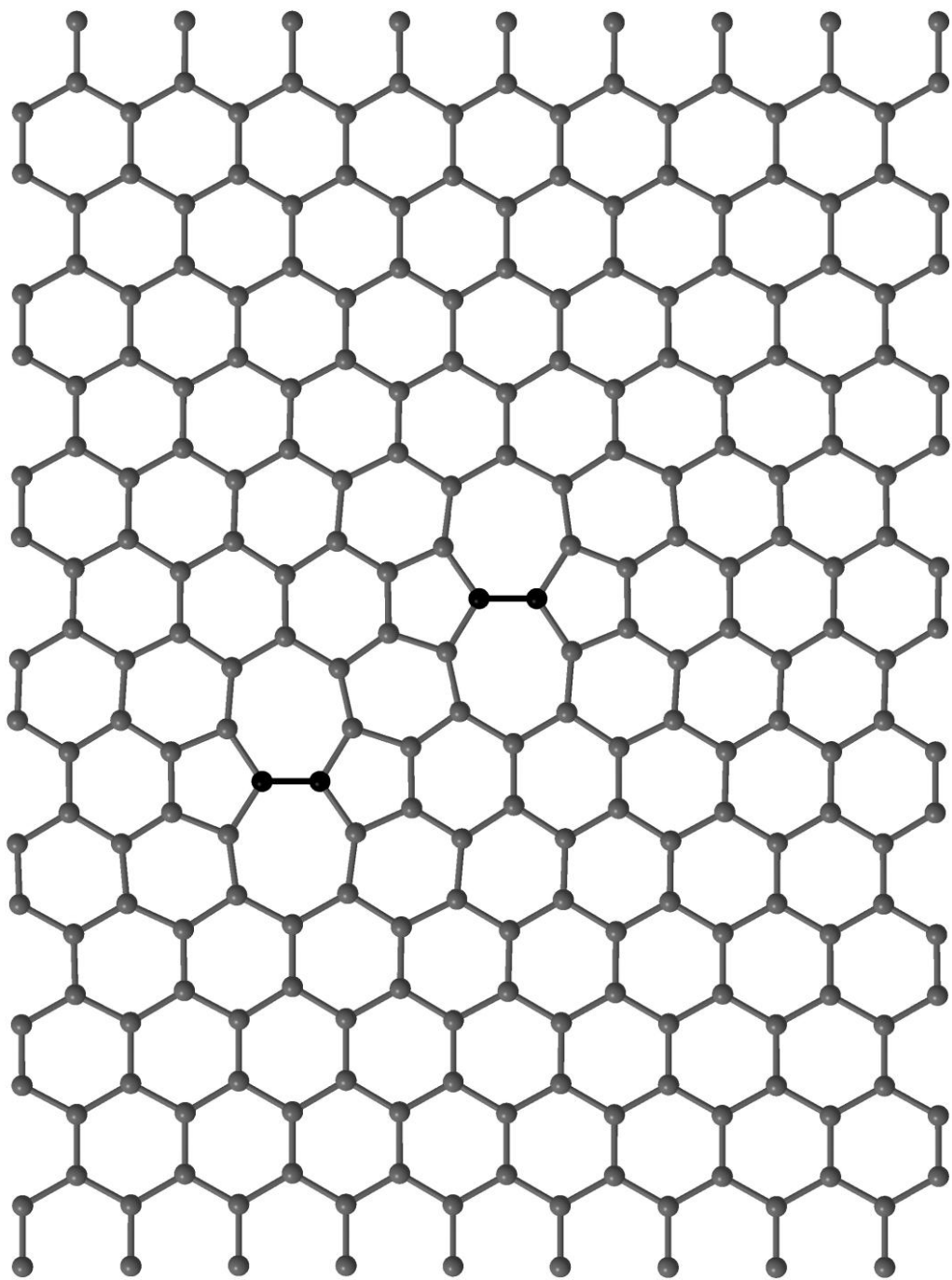

(b)

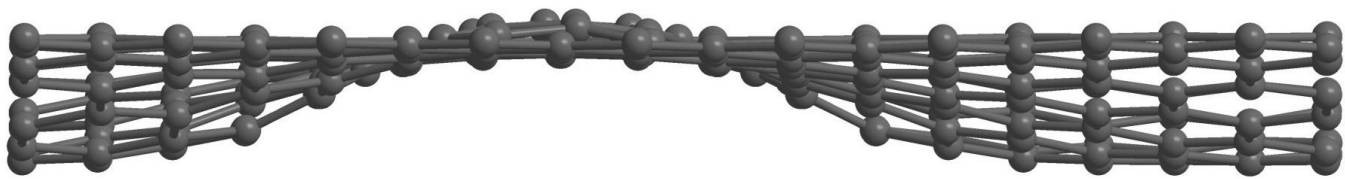

Fig. 3. Same as in Fig. 2 but in configuration II.

We considered five configurations of two Stone-Wales defects corresponding to their different mutual orientations and different (within the dimension of the supercell) distances between them. The structure of each configuration was optimized both in the periods of the supercell and in the coordinates of all constituent atoms. Defects repulse each other (i.e., $E_{int}>0$) in three configurations and attract in two configurations, where this attraction is quite strong. These configurations (I and II) are shown in Figs. 2a and 3a. Both configurations are stable (imaginary frequencies are absent in the spectra of normal oscillations calculated by diagonalizing the Hessian and the maximum frequencies are 1692 and 1675 cm$^{-1}$ for configurations I and II, respectively, which are ≈ 120 cm$^{-1}$ larger than those in the defectless supercell). The cores of defects in configuration I (Fig. 2a) lie on a single straight line. The energy of this configuration is minimal when both defects are sinelike (Fig. 1b), while the nearest atoms of different cores are displaced (perpendicularly to the monolayer) in one direction and, thereby, transverse displacements of atoms in the region between the cores are summed, forming a pronounced maximum with a height of 1.7 Å (Fig. 2b). On the other side of each of the cores, atoms are displaced in the opposite direction. As a result, the maximum transverse dimension of the supercell deformed by two defects is $\Delta z = 2.9$ Å, which is almost twice as large as the value for one sinelike Stone-Wales defect [6]. The distance between defects (defined as the distance between the centers of the cores) is $d = 7.5$ Å and the energy of interaction is $E_{int} = -0.60$ eV. A decrease in $d$ to 5 Å at the conservation of the mutual orientation of the defect cores results in an increase in the total energy of the system, which is due to the presence of a pair of adjacent pentagons between the cores that is energetically unfavorable for graphene. As a result, the energy of interaction, remaining negative, decreases in absolute value ($E_{int} = -$ 0.16 eV). We note that, although $E_{int}$ is negative, defects repulse each other because an increase in $d$ leads to a decrease in $E_{int}$. At the minimum possible value $d = 3$ Å, repulsion is enhanced ($E_{int} = 1.36$ eV) because of the formation of a more unfavorable tetragon.

Another configuration, where defects repulse each other, is shown in Fig. 4. In this configuration, the cores of the defects lie on parallel straight lines, which are perpendicular to the straight segment connecting the centers of the cores. The distance between the defects is $d = 8.5$ Å. The energy of the interaction is $E_{int} = 0.75$ eV.

In configuration II (Fig. 3a), the cores of the defects lie on parallel straight lines, which make an angle with the straight segment connecting the centers of the cores. The energy is minimal when all four atoms of two cores are displaced in one direction from the plane of the monolayer and, in contrast to the cosinelike Stone-Wales defect (Fig. 1c), transverse displacements of two atoms of each core are significantly different (0.04 and 0.57 Å). As in configuration I, the displacements of the nearest atoms of different cores are maximal. Therefore, the profile of transverse displacements (Fig. 3b) has a similar shape, although the value $\Delta z = 2.2$ Å is slightly smaller. The distance between the defects is $d = 6.7$ Å. The energy of the interaction is $E_{int} = -1.24$ eV.

Since $E_{int} \to 0$ at $d \to \infty$, it follows from the results that the equilibrium distance $d_0$ at which the energy of interaction is negative and maximal in absolute value should exist. Apparently, $d_0 \approx 10$ Å or somewhat larger, i.e., on the order of the characteristic dimension of the region of strong distortion of the monolayer by one sinelike or cosinelike Stone-Wales defect (see Fig. 1). To determine $d_0$ more accurately, it is necessary to calculate the energy of interaction between the defects in much larger supercells.

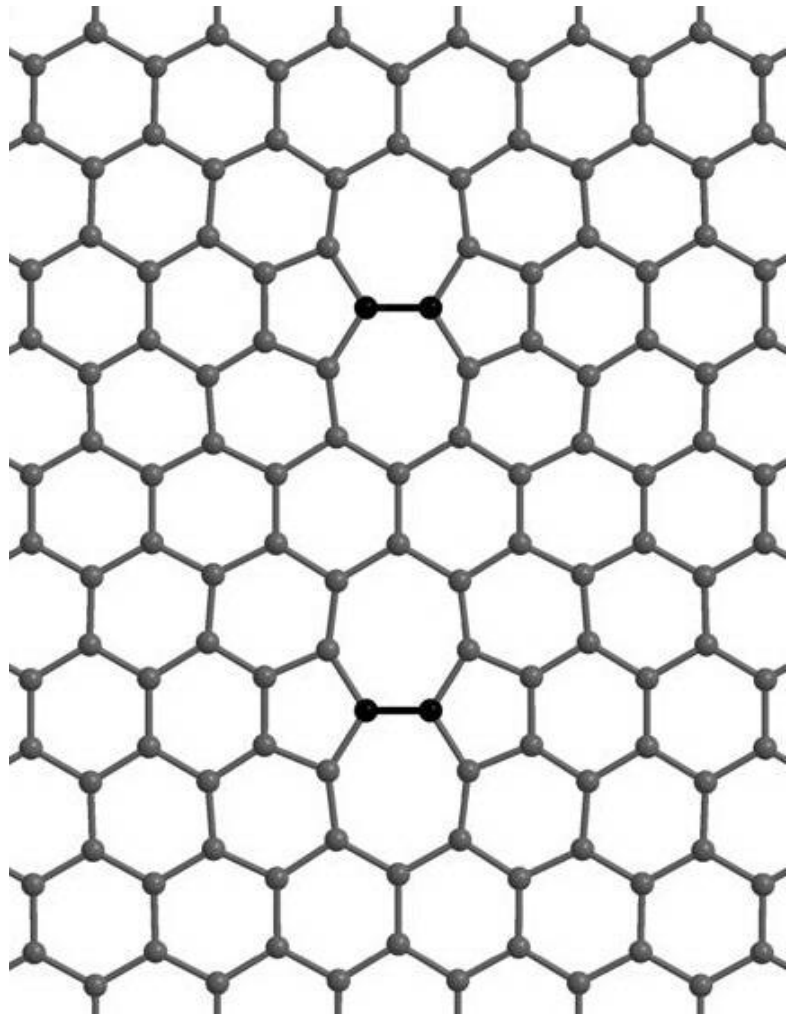

Fig. 4. Configuration in which two Stone-Wales defects repulse each other. The cores of defects are shown in black. Plan view.

The repulsion of a pair of Stone–Wales defects from its images in neighboring supercells makes a positive contribution to the found values of the energy of interaction between two defects $E_{int}$. Consequently, it can be expected that, with an increase in the dimension of the supercell (i.e., with an increase in the distance from defects to images and, correspondingly, with weakening of repulsion), the energy $E_{int}$ not only remains negative but also increases in absolute value.

The attraction between two Stone-Wales defects indicates that these defects may form clusters, although the final conclusion requires analysis of the possibility (and conditions) of their migration over the monolayer. The strong (with an amplitude of several angstroms) wavy transverse deformation of the monolayer by such clusters, as well as by pairs of neighboring Stone-Wales defects, can result in the formation of the experimentally observed "crumpled" texture of graphene [12].

We think that the attraction between Stone-Wales defects in graphene is due to the anisotropic character of deformation of the monolayer by defects. At constructive interference of transverse displacement waves from the nearby defects (Figs. 2b and 3b), the dimension of the deformed region is smaller than the total dimension of the regions of deformation of the monolayer by spaced defects. For this reason, an increase in the energy is smaller. This situation has an analogy with interlayer interstitial (or foreign atoms) in graphite, which attract each other owing to the deformation of the nearest graphite layers if they do not form covalent bonds with these layers [13]. The difference is that interlayer interstitials in graphite are located between deformed layers, whereas Stone-Wales defects in graphene deform layers in which they are located.

In conclusion, we note that the strong bend of fragments of graphene in the presence of Stone-Wales defects in them can promote the formation of various nonplanar carbon nanostructures such as fullerenes and nanotubes.

We are grateful to M.M. Maslov for assistance in the work and discussion of results. The work was performed at National Research Nuclear University MEPhI and was supported by the Russian Science Foundation (project no. 14-22-00098).

# REFERENCES

1. K.S. Novoselov, A.K. Geim, S.V. Morozov, D. Jiang, Y. Zhang, S.V. Dubonos, I.V. Grigorieva, and A.A. Firsov, Science 306, 666 (2004).

2. A.J. Stone and D.J. Wales, Chem. Phys. Lett. 128, 501 (1986).

3. F. Banhart, J. Kotakoski, and A.V. Krasheninnikov, ACS Nano 5, 26 (2011).

4. X. Peng and R. Ahuja, Nano Lett. 8, 4464 (2008).

5. L. Chen, H. Hu, Yu. Quyang, H.Z. Pan, Y.Y. Sun, and F. Liu , Carbon 49, 3356 (2011).

6. J. Ma, D. Alfe, A. Michaelides, and E. Wang, Phys. Rev. B 80, 033407 (2009).

7. A.I. Podlivaev and L.A. Openov, Phys. Solid State 57, 820 (2015).

8. S.N. Shirodkar and U.V. Waghmare, Phys. Rev. B 86, 165401 (2012).

9. V.F. Elesin, V.A. Kashurnikov, L.A. Openov, and A.I. Podlivaev, Sov. Phys. JETP 72, 133 (1991).

10. M.M. Maslov, A.I. Podlivaev, and L.A. Openov, Phys. Lett. A 373, 1653 (2009).

11. L.A. Openov, A.I. Podlivaev, and M.M. Maslov, Phys. Lett. A 376, 3146 (2012).

12. J.C. Meyer, A.K. Geim, M.I. Katsnelson, K.S. Novoselov, T.J. Booth, and S. Roth, Nature  446, 60 (2007).

13. V.F. Elesin and L.A. Openov, Surf. Sci.  442, 131 (1999).
11